\documentclass[twocolumn,twocolappendix,trackchanges]{aastex6}
\RequirePackage{lineno}
\usepackage{amsmath}
\usepackage{amssymb}
\usepackage{amsthm}
\usepackage{natbib,aasdefs,url,bm}
\usepackage{array}
\usepackage{float}
\usepackage{graphicx}
\usepackage{subfigure}
\usepackage{color}
\usepackage{hyperref}
\hypersetup{urlcolor=blue,citecolor=blue,linkcolor=blue}

\newcounter{ichi}
\newcounter{ni}
\newcounter{san}
\newcounter{yon}
\def\be{\begin{equation}}
\def\ee{\end{equation}}
\def\ba{\begin{eqnarray}}
\def\ea{\end{eqnarray}}

\def\aap{\ {A\&A}\ }

\def\apjl{\ {ApJL}\ }
\def\apjs{\ {ApJS}\ }

\shorttitle{Multimessenger Connection in NGC 4151, NGC 4945 and Circinus Galaxy}
\shortauthors{Murase et al.}

\begin{document}

\title{Sub-GeV Gamma Rays from Nearby Seyfert Galaxies and Implications for Coronal Neutrino Emission}


\author{Kohta Murase\altaffilmark{1,2,3}}

\author{Christopher M. Karwin\altaffilmark{4}}

\author{Shigeo S. Kimura\altaffilmark{5}}

\author{Marco Ajello\altaffilmark{6}}

\author{Sara Buson\altaffilmark{7}}


\altaffiltext{1}{Department of Physics; Department of Astronomy \&
Astrophysics; Center for Multimessenger Astrophysics, Institute for Gravitation and the Cosmos, The Pennsylvania State University, University Park, PA 16802, USA}
\altaffiltext{2}{School of Natural Sciences, Institute for Advanced Study, Princeton, NJ 08540, USA}
\altaffiltext{3}{Center for Gravitational Physics and Quantum Information, Yukawa Institute for Theoretical Physics, Kyoto University, Kyoto, Kyoto 606-8502, Japan}
\altaffiltext{4}{NASA Postdoctoral Program Fellow, NASA Goddard Space Flight Center, Greenbelt, MD 20771, USA}
\altaffiltext{5}{Frontier Research Institute for Interdisciplinary Sciences; Astronomical Institute, Graduate School of Science, Tohoku University, Miyagi 980-8578, Japan}
\altaffiltext{6}{Department of Physics and Astronomy, Clemson University, Clemson, SC 29634, USA}
\altaffiltext{7}{Fakult\"at f\"ur Physik und Astronomie, Julius-Maximilians-Universit\"at W\"urzburg, W\"urzburg D-97074, Germany}

\begin{abstract}
Recent observations of high-energy neutrinos by IceCube and gamma rays by the {\it Fermi} Large Area Telescope (LAT) and the MAGIC telescope have suggested that neutrinos are produced in gamma-ray opaque environments in the vicinity of supermassive black holes. In this work, we present 20\,MeV -- 1\,TeV spectra of three Seyfert galaxies whose nuclei are predicted to be active in neutrinos, NGC 4151, NGC 4945 and the Circinus galaxy, using 14.4\,yr of the {\it Fermi} LAT data. 
In particular, we find evidence of sub-GeV excess emission that can be attributed to gamma rays from NGC 4945, as was also seen in NGC 1068. These spectral features are consistent with predictions of the magnetically powered corona model, and we argue that NGC 4945 is among the brightest neutrino active galaxies detectable for KM3Net and Baikal-GVD. On the other hand, in contrast to other reported results, we do not detect gamma rays from NGC 4151, which constrains neutrino emission from the accretion shock model. Future neutrino detectors such as IceCube-Gen2 and MeV gamma-ray telescopes such as AMEGO-X will be crucial for discriminating among the theoretical models. 
\end{abstract}


\section{Introduction}
Since the IceCube Collaboration announced the first evidence for high-energy cosmic neutrinos in 2013 \citep{Aartsen:2013bka,Aartsen:2013jdh}, the detection of many astrophysical TeV--PeV neutrinos has been reported. 
They are produced by hadronuclear ($pp$) and photohadronic ($p\gamma$) interactions, inevitably producing a similar amount of gamma rays. These gamma rays can be absorbed by the Breit-Wheeler process ($\gamma+\gamma\rightarrow e^++e^-$) and reprocessed to lower-energy gamma rays, which can contribute to the cosmic gamma-ray background measured by {\it Fermi} Large Area Telescope \citep[LAT;][]{Ackermann:2014usa}. If all the neutrino sources were gamma-ray transparent, gamma rays accompanied by the large all-sky neutrino flux at neutrino energies of $E_\nu\sim1-10$~TeV \citep{Aartsen:2015ita,IceCube:2020acn} would overshoot the observed diffuse isotropic gamma-ray background. 
This necessitates the existence of hidden cosmic-ray (CR) accelerators that are opaque to GeV-TeV gamma rays, and the vicinity of supermassive black holes (SMBHs) is among the most promising sites \citep{Murase:2015xka}. 

The existence of hidden neutrino sources has been further supported by the recent IceCube observation of neutrino signals from the nearby Seyfert galaxy, \mbox{NGC 1068} \citep{IceCube:2019cia,IceCube2022NGC1068}. The GeV-TeV gamma-ray fluxes seen by LAT \citep{Fermi-LAT:2019yla,Ajello:2023hkh} and MAGIC upper limits \citep[ULs;][]{MAGIC:2019fvw} are much lower than the TeV neutrino flux measured with IceCube. The region has to be compact for the system to be opaque to GeV gamma rays~\citep{Murase:2022dog}, as expected in disk-corona models of active galactic nuclei \citep[AGNs;][]{Murase:2019vdl,Kheirandish:2021wkm}.
This is the main consequence of the GeV-TeV gamma-ray and TeV--PeV neutrino observations. It has been predicted that X-ray bright AGNs are promising targets for neutrino observations \citep[see][as a review]{Murase:2022feu}, while TeV gamma rays are cascaded down to sub-GeV energies, which will be ideal targets for MeV--GeV gamma-ray telescopes. 

In this work, we investigate the multimessenger implications of three nearby Seyfert galaxies, \mbox{NGC 4151}, \mbox{NGC 4945}, and the Circinus galaxy, using {\it Fermi} LAT data. We analyze 14.4\,yr of {\it Fermi} LAT data to obtain the gamma-ray fluxes from these galaxies. Then, we compare the sub-GeV gamma-ray data with CR-induced electromagnetic cascade emission and discuss corresponding neutrino fluxes. 
We use $Q_x=Q/10^{x}$ in cgs units and assume cosmological parameters with $\Omega_m=0.3$, $\Omega_\Lambda=0.7$, and $h=0.7$.

\section{Neutrino and Gamma-Ray Emission from the Hearts of AGNs}
Multimessenger data from \mbox{NGC 1068} suggest that high-energy neutrinos from Seyfert galaxies are likely to originate from compact regions with a size of $R\lesssim30-100~R_S$ \citep{Murase:2022dog}, and X-ray emitting coronae or possible shock regions, located at the vicinity of SMBHs, are the most plausible site of the neutrino production \citep{Murase:2019vdl,Eichmann:2022lxh}. In this work, we consider neutrinos and gamma rays from such coronal regions with $R=10-30~R_S$. 

The spectral energy distribution (SED) of AGNs is modeled empirically \citep{Murase:2019vdl} based on the parameterization as a function of the SMBH mass $M_{\rm BH}$ and the intrinsic X-ray luminosity $L_X$. We use $M_{\rm BH}=1.0\times10^7~M_\odot$ and $L_X=2.6\times10^{42}~{\rm erg}~{\rm s}^{-1}$ for \mbox{NGC 4151} \citep{Bentz:2022aaa,Koss:2022lkb}, $M_{\rm BH}=1.4\times10^6~M_\odot$ and $L_X=3.3\times10^{42}~{\rm erg}~{\rm s}^{-1}$ for \mbox{NGC 4945} \citep{Greenhill:1997kv,Puccetti:2014pqa}, and $M_{\rm BH}=1.7\times10^6~M_\odot$ and $L_X=10^{42.5}~{\rm erg}~{\rm s}^{-1}$ for the Circinus galaxy \citep{Greenhill:2003bz,Tanimoto:2019tbz}, respectively. 
We checked that the adopted SED model \citep[see Figure~2 of][]{Murase:2019vdl} agrees with the observed SEDs of the AGNs considered in this work.

High-energy neutrinos are produced through meson production by $p\gamma$ and $pp$ interactions. We calculate neutrino and gamma-ray emission, considering CR-induced cascades, with the AGN module of the {\sc AMES} \citep[Astrophysical Multimessenger Emission Simulator; e.g.,][]{Murase:2017pfe,Zhang:2023ewt}. 
Details of our steady-state corona model are described in \cite{Murase:2019vdl}, but we use the detailed cross sections of the photomeson production and Bethe-Heitler pair production processes as in \cite{Murase:2022dog}. The model predicts that TeV neutrinos must be accompanied by sub-GeV gamma rays, predicting gamma-ray and neutrino energy fluxes of $E_\gamma F_{E_\gamma}\sim (0.1-1)E_\nu F_{E_{\nu}}$ within 1 order of magnitude. This is a consequence of the multimessenger connection, which is insensitive to model parameters.

\subsection{Magnetically Powered Coronae}
Magnetohydrodynamic (MHD) simulations for accretion flows revealed that the magnetorotational instability (MRI) and subsequent dynamos amplify the magnetic field and develop strong turbulence in accretion flows \citep[e.g.,][]{1998RvMP...70....1B,Kimura:2018clk}. It is believed that magnetic dissipation in the accretion flows heat the plasma and form a hot region called a corona. The observed X-ray emission with a power-law spectrum originates from the Comptonization of disk photons by thermal or bulk electrons.  

The MHD turbulence scatters CRs, and the scattered particles randomly change their energies, and effectively gain energy from the turbulence. 
In the coronal region of AGNs, Alfv\'en and turbulent velocities are mildly relativistic, enabling the turbulence to accelerate CRs up to TeV-PeV energies. In this process, the resulting CR spectrum can be hard with a spectral index of $s\lesssim2$ (for $dN_{\rm CR}/d\varepsilon_p\propto \varepsilon_p^{-s}$, where $\varepsilon_p$ is the CR proton energy in the source frame), and the maximum CR energy typically lies in the range of $\sim0.1-1$~PeV with a reasonable acceleration efficiency ($\eta_{\rm tur}\sim10-100$). The stochastic acceleration by large-scale turbulence has been studied by test-particle simulations with MHD \citep{Kimura:2016fjx,Kimura:2018clk,Sun:2021ods} and large-scale particle-in-cell (PIC) simulations \citep{Zhdankin:2018lhq,Comisso:2019frj,Comisso:2022iqy}. 
Alternatively, magnetic reconnections accelerate nonthermal particles efficiently, especially in the relativistic regime \citep[$\sigma_{\rm mag}=B^2/(4\pi n mc^2)>1$;][]{Hoshino:2012aaa,Guo:2020fni}.
PIC simulations have demonstrated that the development of MRI turbulence triggers magnetic reconnections, leading to nonthermal particle production \citep[e.g.,][]{Hoshino:2015cka,Kunz:2016reh,Bacchini:2022myw}. \cite{Kheirandish:2021wkm} suggested such a magnetic reconnection model to explain the neutrino data of \mbox{NGC 1068} \citep[see also][]{Mbarek:2023yeq,Fiorillo:2023dts}. 
 
In this work, we consider the magnetically powered corona model with stochastic acceleration \citep{Murase:2019vdl}, where we solve the Fokker-Planck equation to obtain the CR spectrum and calculate the resulting neutrino and gamma-ray spectra taking into account electromagnetic cascades through $\gamma\gamma\rightarrow e^+e^-$, synchrotron and inverse-Compton emission processes. The coronal plasma is assumed to be geometrically thick and advected to an SMBH with the infall velocity $V_{\rm fall}=\alpha\sqrt{GM_{\rm BH}/R}$, where $\alpha=0.1$ is the viscosity parameter. The magnetic field is obtained through the plasma beta set to $\beta=1$, where synchrotron cascade emission is important for magnetized coronae with $\beta\lesssim1$ and the cascaded $\sim0.1–10$~MeV gamma-ray flux is a robust signature \citep{Murase:2019vdl}. Then, the remaining principal parameters of the model are for CR inputs regarding the CR normalization and maximum energy. We consider two specific cases. For model A we assume the CR injection efficiency\footnote{For \mbox{NGC 1068}, the corresponding CR pressure is 50\% of the thermal pressure with the virial temperature for $L_X=3\times{10}^{43}~{\rm erg}~{\rm s}^{-1}$ and $d=10$~Mpc \citep{Ajello:2023hkh}.} and $\eta_{\rm tur}=70$ that are used to explain the IceCube data of \mbox{NGC 1068} \citep[see Figure~4 of][]{IceCube2022NGC1068} with $R=30~R_S$. 
For model B, the CR pressure $P_{\rm CR}$ is set to 8\% of the thermal pressure with the virial temperature ($P_{\rm vir}$). Because the coronal pressure would be dominated by magnetic fields, this assumption is reasonable. The emission radius is assumed to be $R=10~R_S$, in which more neutrinos and gamma rays may be produced due to $p\gamma$ interactions. For the three AGNs studied in this work, model B gives more optimistic predictions for the neutrino and gamma-ray fluxes.


\subsection{Accretion Shocks}
Accretion shocks have been discussed as a CR acceleration site for a long time \citep[e.g.,][]{Stecker:1991vm,Inoue:2019yfs,Anchordoqui:2021vms}. 
In this model, a spherical accretion flow, aside from the accretion disk flow, is assumed to form a shock with a velocity of $V_s\approx V_{\rm ff}=\sqrt{GM_{\rm BH}/R}$
in the vicinity of the SMBH, and $s\gtrsim2$ is expected based on diffusive shock acceleration theory \citep[e.g.,][]{Drury:1983zz}. However, the neutrino data of \mbox{NGC 1068} suggest that the particle acceleration efficiency, $\eta_{\rm sho}$, has to be much larger than the canonical value in the Bohm limit ($\eta_{\rm sho}\sim1-10$) not to violate the constraint on the neutrino break/cutoff \citep[see Figure~5 of][]{Kheirandish:2021wkm}. Thus, in this work, we {\it ad hoc} fix the proton maximum energy to 100~TeV (implying $\eta_{\rm sho}\gg10$) and consider the largest CR luminosity $L_{\rm CR}$ allowed by the LAT data. Then, we calculate neutrino and gamma-ray spectra assuming the steady-state injection of CRs for given $R$ and SEDs.

In the accretion shock model, magnetic fields are expected to be weaker than the magnetically powered corona model, and we set $B=30$~G throughout this work \citep{Inoue:2023bmy}. For such low magnetizations, the results on gamma-ray spectra are largely insensitive to CR spectra as long as electromagnetic cascades proceed mainly through $\gamma\gamma\rightarrow e^+e^-$ and inverse-Compton emission.   
Note that we only consider CR-induced cascade gamma rays without any primary nonthermal electron component, unlike \cite{Inoue:2023bmy}, who focused on gamma rays from primary electrons only with attenuation (without cascades). The primary electron component would add an additional sub-GeV gamma-ray component, which would result in a lower neutrino flux in order to be consistent with the given sub-GeV gamma-ray data.

\section{Data Analysis}
\label{sec:data_analysis}
\begin{table}[b]
\begin{deluxetable}{lcc}
\tablecaption{Summed Likelihood Components}
\tablehead{
\colhead{Energy [GeV]} & \colhead{Z Max [deg]} & \colhead{PSF Types}    
}
\startdata
$0.02 - 0.1$    &80       &3 \\
$0.1 - 0.3$      &90       &2, 3 \\
$0.3 - 1$      &100       &1, 2, 3 \\
$1 - 1000$      &105       &0, 1, 2, 3 
\enddata
\label{tab:data}
\tablecomments{These energy bins distinguish different components of the joint likelihood analysis; however, the underlying fits still use eight energy bins per decade, as described in the text. Z Max specifies the maximum zenith angle which controls the Earth limb contamination.}
\end{deluxetable}
\end{table}

\begin{figure*}[t]
\begin{center}
\includegraphics[width=0.45\linewidth]{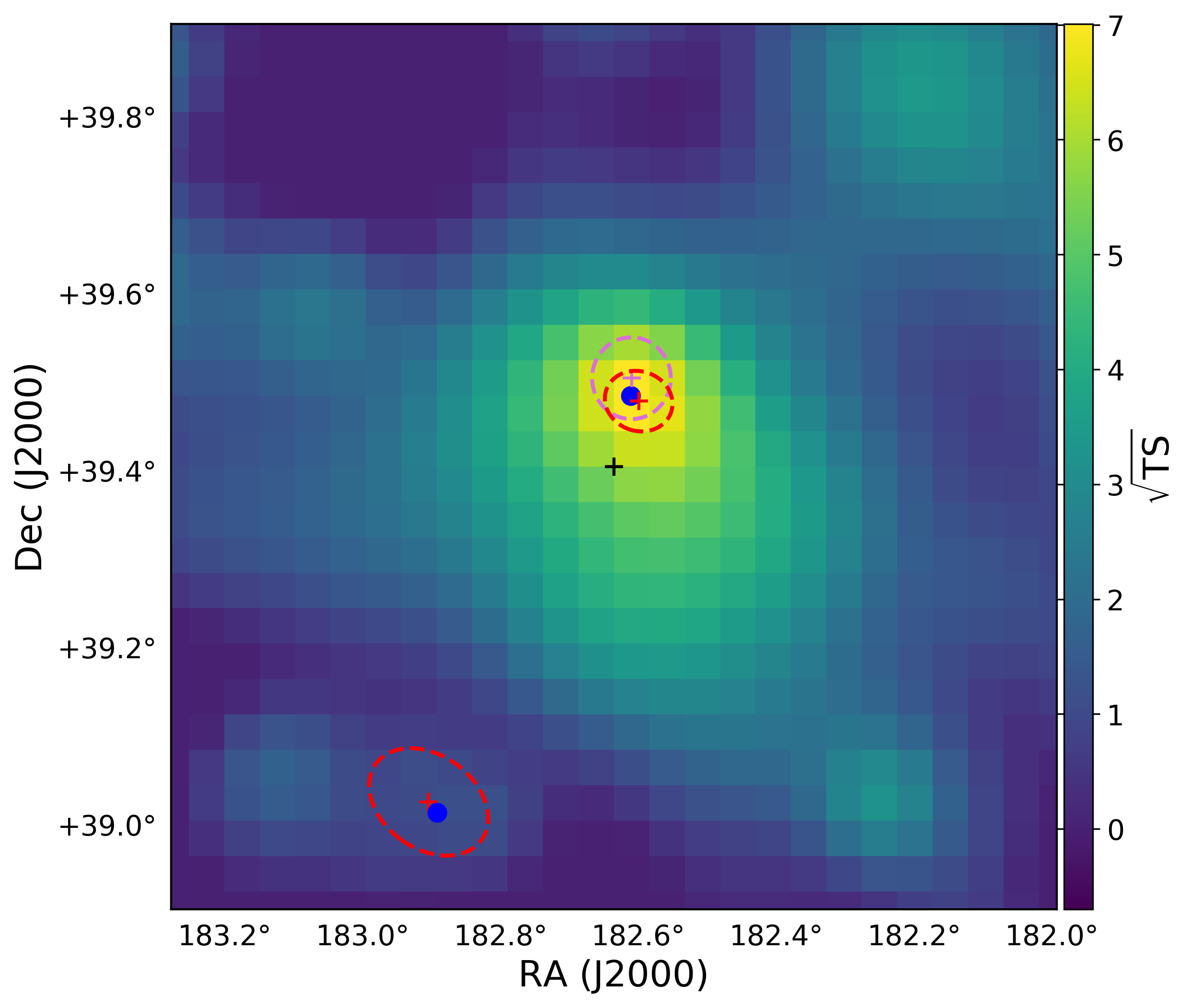}
\includegraphics[width=0.45\linewidth]{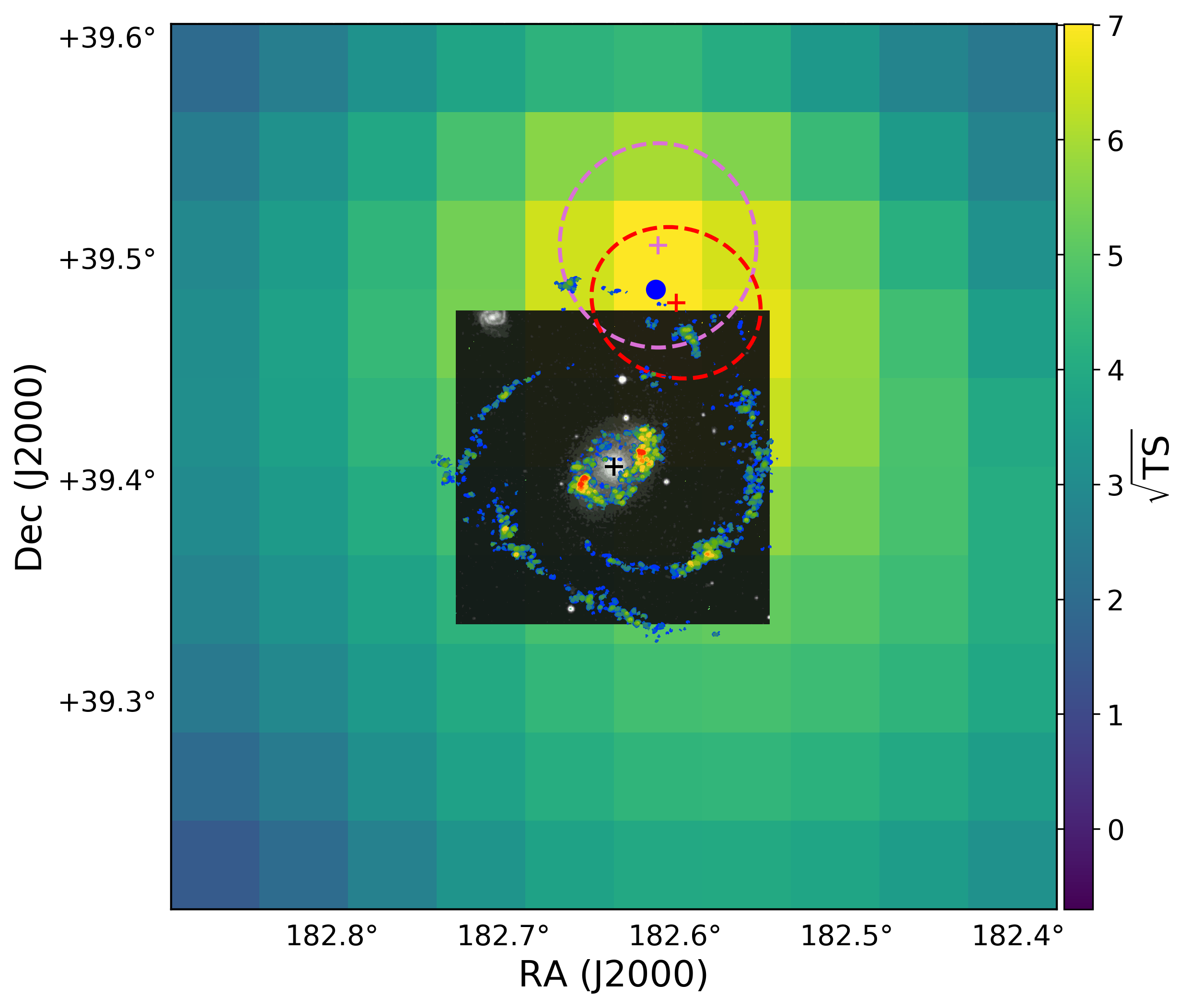}
\caption{Left: TS map of the \mbox{NGC 4151} field. Contours show the 95\% localization uncertainty. Red contours are from the 4FGL-DR3 catalog, and the pink contour is calculated based on the peak in the TS map. Blue circles show the locations of the BL Lacs that are associated with the 4FGL-DR3 sources (\mbox{4FGL 1211.6+3901} towards the lower left, $0.43^\circ$ separation, and 4FGL 1210.3+3928 closest to \mbox{NGC 4151}, $0.08^\circ$ separation). Right: Same as the left plot but zoomed in on \mbox{NGC 4151}. Overlaid are multiwavelength images of \mbox{NGC 4151} in both optical (black and white image) and radio (H I, shown with the rainbow color map).}
\label{fig:localization}
\end{center}
\end{figure*}

We study three nearby Seyfert galaxies, \mbox{NGC 4151}, \mbox{NGC 4945}, and the Circinus galaxy. These objects were all reported as LAT sources and are also very bright in the hard X-ray band. In the magnetically powered corona model, the neutrino luminosity is roughly proportional to the X-ray luminosity \citep{Murase:2019vdl,Kheirandish:2021wkm}, and thus \mbox{NGC 1068} and \mbox{NGC 4151} (\mbox{NGC 4945} and Circinus) are predicted to be among the brightest sources in the northern (southern) neutrino sky among Seyfert galaxies in the {\it Swift} BASS catalog with LAT detection\footnote{With the {\it Swift} BASS catalog considering the 0.2-10~keV and 14-195~keV bands for the intrinsic X-ray flux, the brightest Seyfert galaxies in the northern sky, including the near horizon, are \mbox{NGC 1068} and \mbox{NGC 4151}, and \mbox{NGC 1068} is expected to be the most promising neutrino source, especially with the {\it NuSTAR} flux. In the southern sky, \mbox{NGC 4945}, \mbox{ESO 138-1}, and Circinus are the brightest. We focus on \mbox{NGC 4945} and Circinus, which were reported as LAT sources, and the high luminosity of \mbox{ESO 138-1} would suppress coronal gamma-ray emission more severely.}. Prior to this work, the results of \mbox{NGC 1068} (being the brightest in the IceCube sky) are presented in \cite{Ajello:2023hkh}. 

We analyze data collected by the LAT~\citep{Fermi-LAT09,Fermi-LAT13,Bruel+18} from 2008 August 04 to 2023 January 5 (14.4 yr). We select an energy range of \mbox{20~MeV}--\mbox{1~TeV}, and bin the data using eight energy bins per decade. The pixel size is $0.08^\circ$. We use a $10^\circ \times 10^\circ$ region of interest (ROI) centered on each respective galaxy. The standard data filters are used: DATA\_QUAL$>$0 and LAT\_CONFIG==1. The analysis is performed using Fermipy (v1.2)\footnote{Available at \url{https://fermipy.readthedocs.io/en/latest/}}, which utilizes the underlying Fermitools (v2.2.0). 

We select photons corresponding to the P8R3\_SOURCE\_V3 instrument response. In order to optimize the sensitivity of our analysis, we implement a joint likelihood fit with the four point spread function (PSF) event types available in the Pass 8 data set\footnote{For more information on the different PSF types see \url{https://fermi.gsfc.nasa.gov/ssc/data/analysis/documentation/Cicerone/Cicerone_Data/LAT_DP.html}.}. The data are divided into quartiles corresponding to the quality of the reconstructed direction, from the lowest quality quartile (PSF0) to the best quality quartile (PSF3). Each subselection has its own binned likelihood instance that is combined in a summed likelihood function for the ROI. This is easily implemented in Fermipy by specifying the components section in the configuration file. We include different event types depending on the energy interval. This is motivated by the energy-dependence of the LAT instrument response. Additionally, in order to reduce the contamination from the Earth's limb, we apply an energy-dependent cut on the maximum zenith angle. These selections are similar to those used for the 4FGL-DR3 catalog \citep{Fermi-LAT:2022byn}, and they are summarized in Table~\ref{tab:data}. Note, however, that the 4FGL-DR3 did not use photon energies below \mbox{50~MeV}. Each PSF type also has its own corresponding isotropic spectrum, namely, iso\_P8R3\_SOURCE\_V3\_PSF{\it i}\_v1, for $i$ ranging from 0 to 3. The isotropic spectrum file is defined from \mbox{34~MeV} to \mbox{878~GeV}, and so we use a power-law extrapolation for energies outside this range. The Galactic diffuse emission is modeled using the standard component (gll\_iem\_v07). It is only defined between \mbox{50~MeV} and \mbox{814~GeV}, so we likewise use a power-law extrapolation. The point source emission is modeled using the 4FGL-DR3 catalog \citep[gll\_psc\_v28;][]{Fermi-LAT:2022byn}. In order to account for photon leakage from sources outside of the ROI due to the PSF of the detector, the model includes all 4FGL sources within a $15^\circ \times 15^\circ$ region. The energy dispersion correction (edisp\_bins=--1) is enabled for all sources except the isotropic component.

\begin{figure*}[t]
\includegraphics[width=0.33\linewidth]{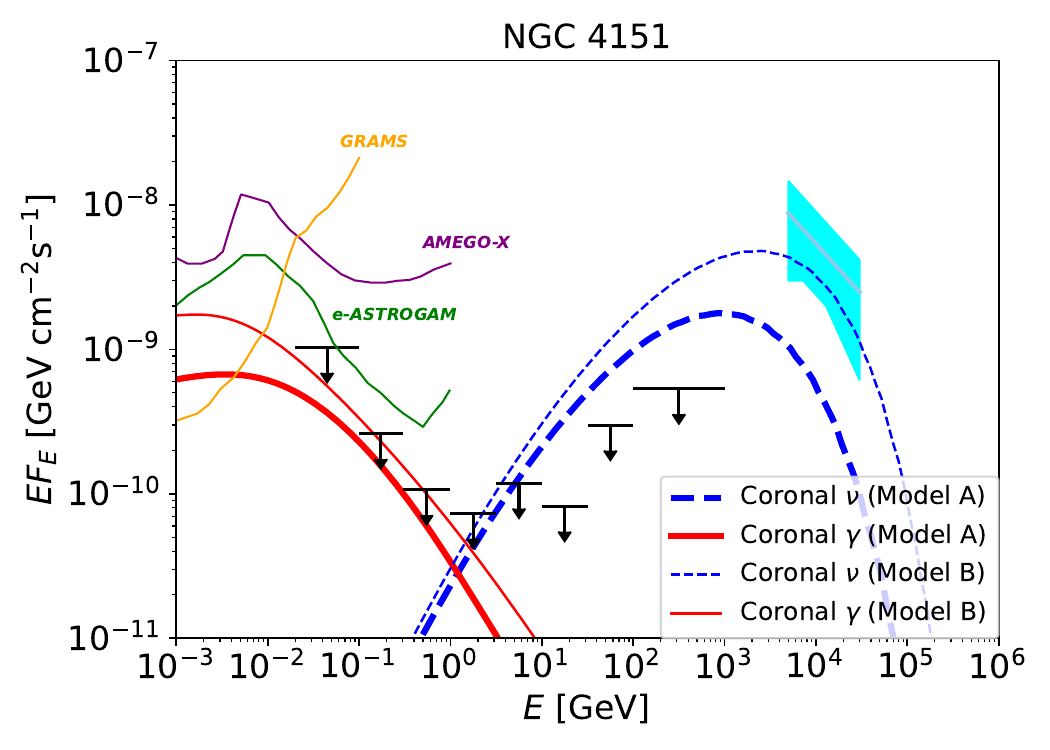}
\includegraphics[width=0.33\linewidth]{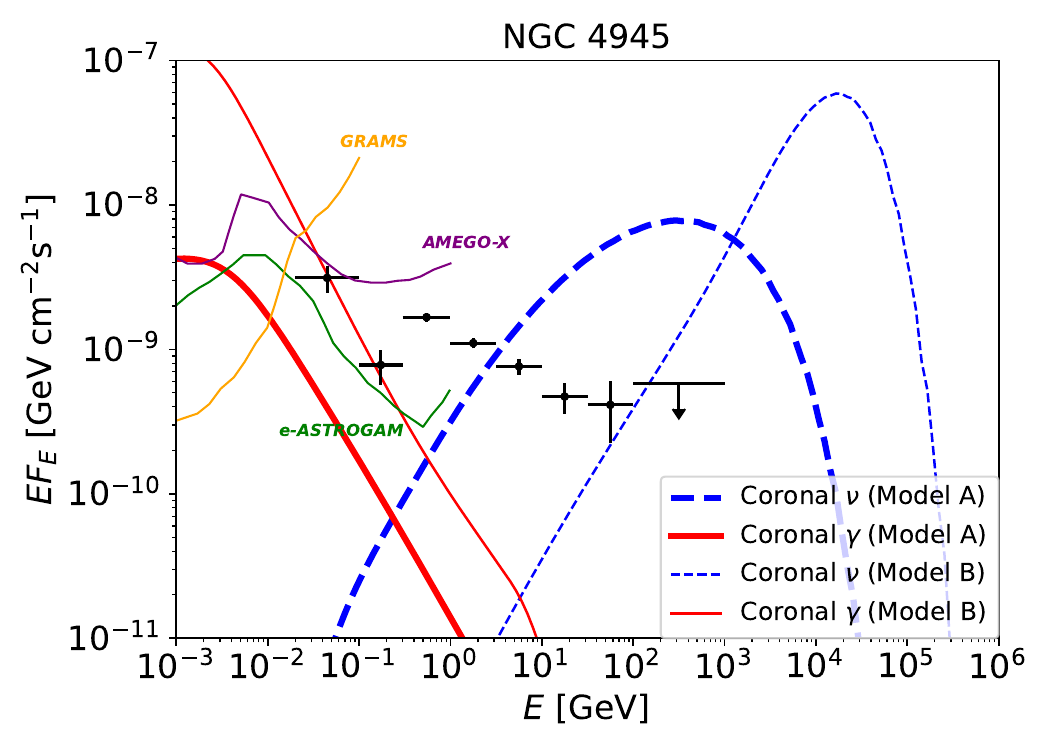}
\includegraphics[width=0.33\linewidth]{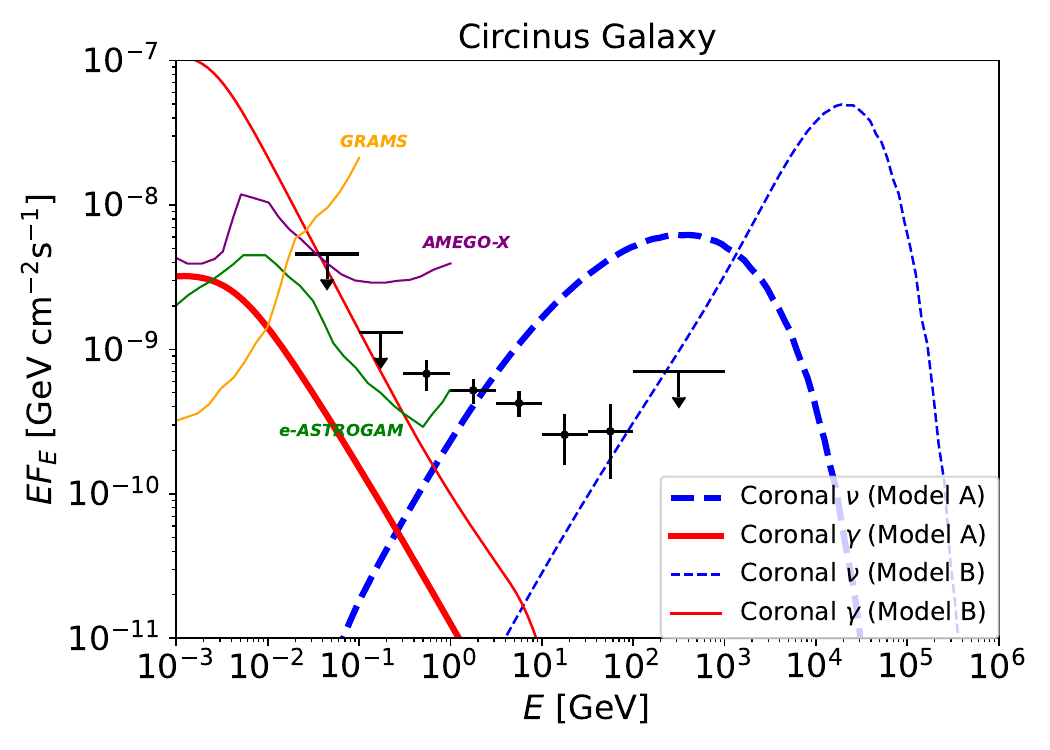}
\caption{
Multimessenger SEDs of \mbox{NGC 4151}, \mbox{NGC 4945} and Circinus galaxy in the magnetically powered corona model. Black points and ULs are {\it Fermi} LAT data in gamma rays (this work). The error bars on the LAT data points are 1$\sigma$, and the ULs are plotted for bins with TS$<$4, and they are shown at the $95\%$ C.L. The sky-blue shaded region represents IceCube data in neutrinos \citep{IceCube:2023nai,TAUP2023}. Sensitivities for e-ASTROGAM \citep{e-ASTROGAM:2016bph} and GRAMS \citep{Aramaki:2019bpi} with an effective exposure time of 1~yr are overlaid, together with the AMEGO-X sensitivity for the 3~yr mission~\citep{Caputo:2022xpx}. For model A, we use parameters to explain \mbox{NGC 1068} as in \cite{Murase:2019vdl}, where the emission radius is set to $R=30R_S$ with $\eta_{\rm tur}=70$. For model B, we assume $R=10R_S$ and the CR pressure is set to $P_{\rm CR}/P_{\rm vir}=8$\%, where $\eta_{\rm tur}=100$ for NGC 4151 and $\eta_{\rm tur}=10$ for \mbox{NGC 4945} and Circinus are used.      
\label{fig:AGN1}
}
\end{figure*}

\begin{figure*}[t]
\includegraphics[width=0.33\linewidth]{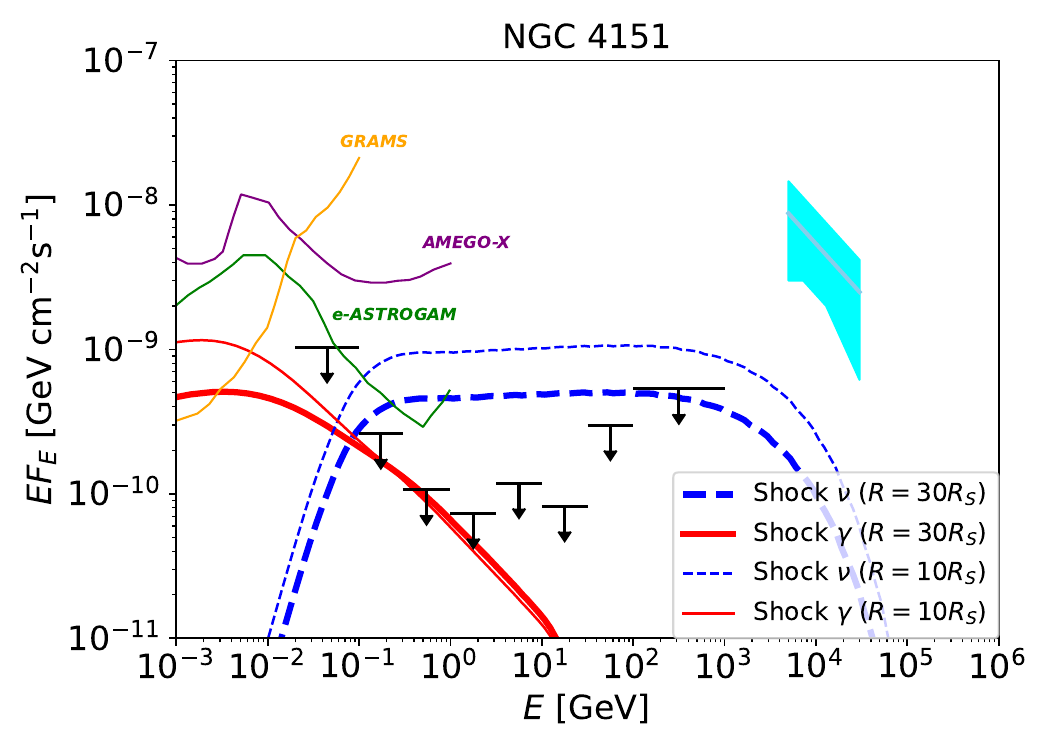}
\includegraphics[width=0.33\linewidth]{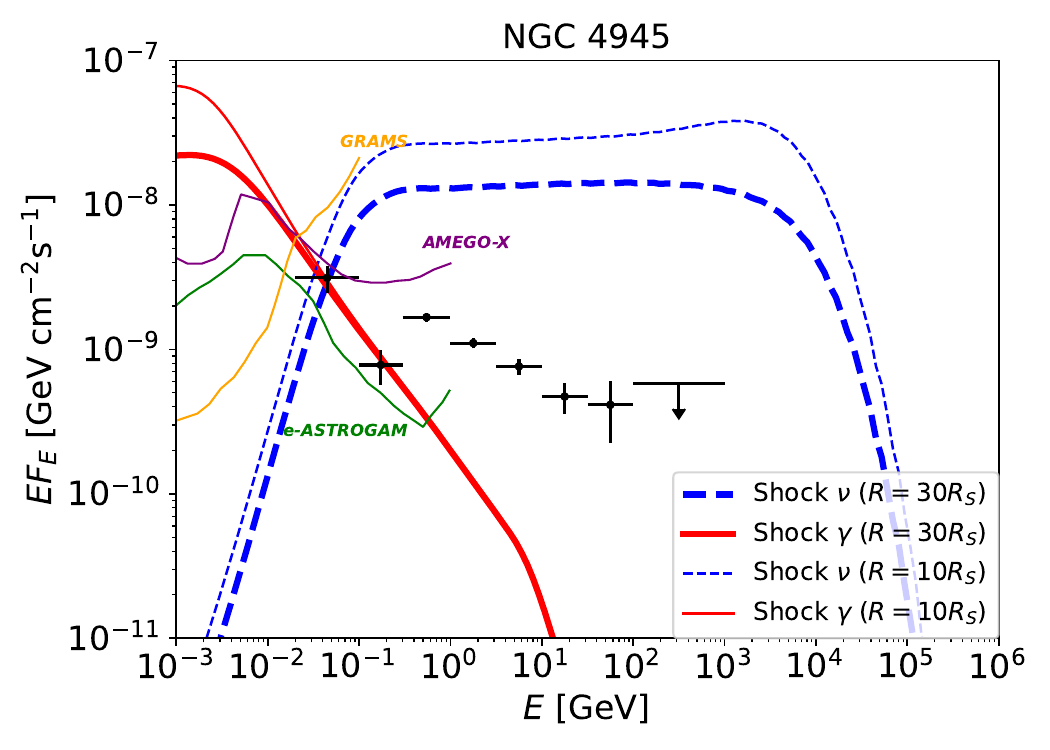}
\includegraphics[width=0.33\linewidth]{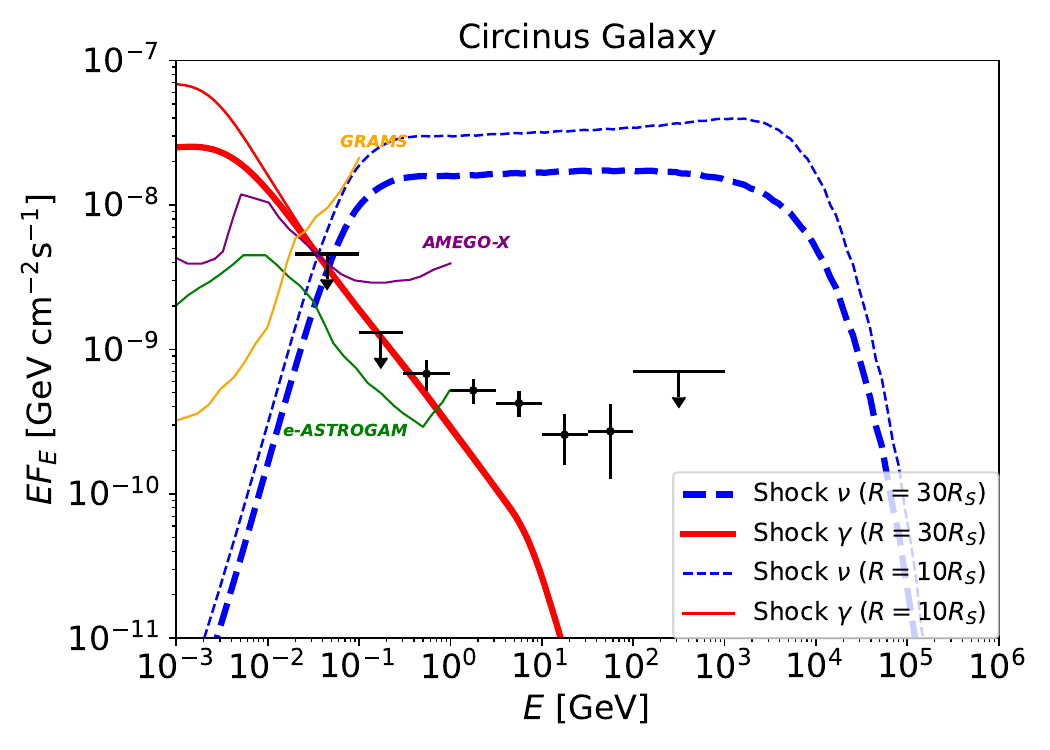}
\caption{
Same as Figure~\ref{fig:AGN1} but in the accretion shock model. The maximum injection luminosities of CRs for $R=30R_S$ are $L_{\rm CR}=3\times10^{42}~{\rm erg}~{\rm s}^{-1}$ (\mbox{NGC 4151}), $L_{\rm CR}=6\times10^{42}~{\rm erg}~{\rm s}^{-1}$ (\mbox{NGC 4945}), and $L_{\rm CR}=10^{43}~{\rm erg}~{\rm s}^{-1}$ (Circinus). For $R=10R_S$, they are $L_{\rm CR}=10^{43}~{\rm erg}~{\rm s}^{-1}$, $L_{\rm CR}=2\times10^{43}~{\rm erg}~{\rm s}^{-1}$, and $L_{\rm CR}=3\times10^{43}~{\rm erg}~{\rm s}^{-1}$, respectively.
\label{fig:AGN2}
}
\end{figure*}

Before calculating SEDs, we first perform an initial fit for each of the three ROIs, in which we optimize the model, freeing the normalization and index of the Galactic diffuse emission, the normalization of the isotropic emission, and all point sources (normalization and index) within 3$^\circ$. We then find new sources using the Fermipy function \textit{find\_sources}, which generates test statistic (TS) maps and identifies new sources based on peaks in the TS. The TS maps are generated using a power-law spectral model with a photon index of $2.0$. The minimum separation between two point sources is set to $0.5^\circ$, and the minimum TS for including a source in the model is set to 16. Finally, we perform a second fit in a similar way as the first, with the exception that we free all point sources within 5$^\circ$ having a TS$>25$.

Among the three Seyfert galaxies, the LAT observations of \mbox{NGC 4151} are complicated by the presence of two nearby BL Lacs, \mbox{4FGL J1211.6+3901} and \mbox{4FGL 1210.3+3928}. In order to determine the most likely origin of the corresponding gamma-ray emission, we relocalize the point source closest to \mbox{NGC 4151}. This is done following a similar procedure as the ROI optimization described above. We use the same data selection, with the exception that we use a pixel size of 0.04$^\circ$. In the first step of the optimization, we only include point sources from the 4FGL-DR3; i.e., we do not include a point source at the location of \mbox{NGC 4151}. Then, before finding new sources, we remove \mbox{4FGL 1210.3+3928} from the model, thus allowing the fitting algorithm to find the best-fit location of the associated source. The resulting localization is shown in Figure~\ref{fig:localization}. The pink contour shows the 95\% localization uncertainty, based on the peak in the TS map. Red contours show the 95\% localization uncertainty from the 4FGL-DR3, and the blue circles show the BL Lacs which are associated with the 4FGL sources. As can be clearly seen, the gamma-ray source near \mbox{NGC 4151} is most likely associated with \mbox{4FGL 1210.3+3928} (and hence the BL Lac), consistent with the 4FGL catalog. We also verified that the results remain consistent when using energies \mbox{$E_\gamma >1$~GeV}. Additionally, we checked with the preliminary version of the 4FGL-DR4, also finding consistent results. We therefore keep \mbox{4FGL 1210.3+3928} in the model when calculating the SED for \mbox{NGC 4151}.

After optimizing the ROIs, SEDs are calculated for each source using the Fermipy SED analysis. This method computes an SED by performing independent fits for the flux normalization in each energy bin. For the calculation, we combine the original energy bins into larger bins, using two bins per decade between \mbox{100~MeV} and \mbox{100~GeV}, as well as a single bin between \mbox{20} and \mbox{100~MeV} and a single bin between \mbox{100} and \mbox{1000~GeV}. The normalization in each bin is fit using a power-law spectral model with a fixed index of 2.0. The parameters of the background components are held fixed at their best-fit values from the baseline fit. Further details of the SED analyses for the three Seyfert galaxies are provided in Appendix~\ref{sec:appendix}. The resulting SEDs of \mbox{NGC 4151}, \mbox{NGC 4945}, and Circinus are shown in Figures~\ref{fig:AGN1} and \ref{fig:AGN2}.

\section{Implications and Discussions}
\label{sec:general}
GeV gamma rays mainly interact with X rays from the corona. The two-photon annihilation optical depth for a photon index of $\Gamma_X\approx2$ is \citep{Murase:2022dog}
\begin{eqnarray}
\tau_{\gamma\gamma}(\varepsilon_\gamma)&\approx&
\eta_{\gamma\gamma}\sigma_{T}R n_X{(\varepsilon_\gamma/\tilde{\varepsilon}_{\gamma\gamma-X})}^{\Gamma_X-1}\nonumber\\
&\simeq&24~{[R/(30R_S)]}^{-1}M_{\rm BH,7}^{-1}\tilde{L}_{X,42.4}(\varepsilon_\gamma/1~\rm GeV),\,\,\,\,\,\,\,\,\,
\label{eq:gg}
\end{eqnarray}
where $\eta_{\gamma\gamma}\sim0.1$ is a numerical coefficient depending on $\Gamma_X$, $\sigma_T\approx6.65\times{10}^{-25}~{\rm cm}^2$, $\tilde{\varepsilon}_{\gamma\gamma-X}=m_e^2c^4/\varepsilon_X\simeq0.26~{\rm GeV}~{(\varepsilon_X/1~{\rm keV})}^{-1}$, $\varepsilon_X$ is the reference X-ray energy, and $n_X\approx \tilde{L}_X/(2\pi R^2 c \varepsilon_X)$ is used. Here $\tilde{L}_X$ is the differential X-ray luminosity. 
The numerical results shown in Figures~\ref{fig:AGN1} and \ref{fig:AGN2} are consistent with Equation~(\ref{eq:gg}), and we may expect \mbox{$\lesssim0.1$~GeV} gamma rays. 
For \mbox{NGC 4945} and Circinus, while GeV emission is detected, its origin should be different, e.g., star-forming activities, as seen in \mbox{NGC 1068} \citep{Ajello:2023hkh}. Note that hadronic emission from the starburst region predicts a low-energy break in the gamma-ray spectrum due to the decay of neutral pions, although there could be leptonic contributions from inverse-Compton and bremsstrahlung emission. 
However, detailed studies of the GeV emission are beyond the scope of this work.  

\mbox{NGC 4151} is known to be a Seyfert 1.5 galaxy, showing the characteristic features of both types 1 (dominated by broad-line components) and 2 (dominated by narrow-line components). NGC 4151 is Compton thin, unlike the other two Seyferts and \mbox{NGC 1068}. The escape of GeV gamma rays is easier than in \mbox{NGC 1068} due to its lower intrinsic X-ray luminosity, which makes the strong limits in the GeV range important.
As shown in the left panels of Figures~\ref{fig:AGN1} and \ref{fig:AGN2}, we find that the magnetically powered corona and accretion shock models are consistent with the ULs obtained in this work. Nevertheless, the data give interesting constraints on neutrino emission. In the magnetically powered corona model, model B predicts a neutrino flux that can explain the possible neutrino excess emission found in the IceCube data~\citep{IceCube:2023nai,TAUP2023,Neronov:2023aks}. In the case of \mbox{NGC 4151}, the effective maximum CR energy\footnote{In the diffusive escape-limited case, the maximum energy found in the CR distribution is effectively $\sim10-30$ higher than the energy inferred from equating the acceleration time with diffusive escape time~\citep{Becker:2006nz,Kimura:2014jba}.} is limited by the diffusive escape rather than the Bethe-Heitler cooling process that is relevant for luminous AGNs, including \mbox{NGC 1068}.
On the other hand, due to the LAT ULs in the GeV range, the neutrino flux is unlikely to be explained in the accretion shock model even if magnetic fields change. This demonstrates that observations of \mbox{NGC 4151}-like galaxies are useful for discriminating between the magnetically powered corona and accretion shock models. When we include not only \mbox{2-10~keV} but also \mbox{14-195~keV} data, the neutrino brightness of \mbox{NGC 4151} is next to \mbox{NGC 1068}, where the ranking is higher than that in \cite{Kheirandish:2021wkm}, and \mbox{NGC 4151} will be an important target for IceCube-Gen2 which can reach a sensitivity of $E_\nu F_{E_{\nu}}\sim{10}^{-9}~{\rm GeV}~{\rm cm}^{-2}~{\rm s}^{-1}$ \citep{IceCube-Gen2:2020qha}, especially in the magnetically powered corona model (see left panel of Figure~\ref{fig:AGN1}). 

For \mbox{NGC 4945}, as shown in the middle panels of Figures~\ref{fig:AGN1} and \ref{fig:AGN2}, we find that GeV and higher-energy emission has a break at \mbox{$E_\gamma\sim1$~GeV}, which is consistent with a pionic gamma-ray component from $\pi^0\rightarrow\gamma\gamma$, as expected in models relying on wind-torus interactions~\citep{Inoue:2022yak} and/or star-forming activities~\citep{Aguilar-Ruiz:2020bzq}. In addition, the LAT data around $\sim30$~MeV shows a clear excess compared to the pionic gamma-ray component. However, it is important to note that the LAT performance below \mbox{50~MeV} quickly deteriorates, and the instrument response at these low energies has not been well characterized. We have performed a number of tests to validate the robustness of the observed low-energy spectral feature in \mbox{NGC 4945} (summarized in Appendix~\ref{sec:appendix}), and none of them have shown any indication of the signal being spurious. But nevertheless, this feature should still be interpreted with caution due to the limitations of the LAT performance below \mbox{50~MeV}.
That said, this signal is very interesting because a similar excess flux over the pionic gamma-ray component was reported for \mbox{NGC 1068}~\citep{Ajello:2023hkh}. More intriguingly, the sub-GeV gamma-ray signatures of these two most promising neutrino-bright AGNs are consistent with predictions of the magnetically powered corona model, and the \mbox{$20-100$~MeV} data of \mbox{NGC 4945} can be explained if the CR pressure is $\sim(5-10)$\% of the thermal pressure with the virial temperature (see Figure~\ref{fig:AGN1} middle). Corresponding to the systematic uncertainty with a factor of $2-3$ in the lowest-energy bin data, theoretical fluxes can be even lower by reducing $P_{\rm CR}$ and/or increasing $\eta_{\rm tur}$.
For Circinus, as shown in the right panels of Figures~\ref{fig:AGN1} and \ref{fig:AGN2}, we only have ULs at sub-GeV energies, which are also consistent with both of the corona and accretion shock models. 
In all the three Seyfert galaxies, the model B fluxes shown in Figure~\ref{fig:AGN1} can be regarded as ULs allowed by the LAT data, while they lie below the COMPTEL sensitivity of $E_\gamma F_{E_{\gamma}}\sim{\rm a~few}\times{10}^{-7}~{\rm GeV}~{\rm cm}^{-2}~{\rm s}^{-1}$~\citep{e-ASTROGAM:2016bph}.

The corona model predicts that \mbox{NGC 4945} and the Circinus galaxy are Seyfert galaxies that are among the most promising for neutrino detection, and the all-flavor neutrino flux reaches $E_\nu F_{E_{\nu}}\sim{10}^{-8}-{10}^{-7}~{\rm GeV}~{\rm cm}^{-2}~{\rm s}^{-1}$ as seen in Figure~\ref{fig:AGN1}. They are promising targets for Northern Hemisphere neutrino detectors such as KM3Net \citep{Adrian-Martinez:2016fdl}, Baikal-GVD \citep{Baikal-GVD:2019kwy}, P-ONE \citep{P-ONE:2020ljt}, and Trident \citep{Ye:2022vbk}. For example, the sensitivity of KM3Net is $E_\nu F_{E_{\nu}}\sim{\rm a~few}\times{10}^{-9}~{\rm GeV}~{\rm cm}^{-2}~{\rm s}^{-1}$. We also encourage IceCube searches for starting tracks and showers in the southern sky\footnote{\cite{IceCube:2019cia} gave $E_\nu F_{E_{\nu}}\lesssim2\times {10}^{-8}~{\rm GeV}~{\rm cm}^{-2}~{\rm s}^{-1}$ for an $E_\nu^{-2}$ spectrum but it is not sensitive to energies below \mbox{$10-30$~TeV} due to the severe muon background.}.

\section{Summary}\label{sec:summary}
We studied gamma-ray emission from three nearby Seyfert galaxies, \mbox{NGC 4151}, \mbox{NGC 4945}, and the Circinus galaxy. The theoretical models predict that the neutrino flux is approximately proportional to the intrinsic X-ray luminosity. \mbox{NGC 1068} and \mbox{NGC 4151} are the most promising neutrino sources in the IceCube sky, while \mbox{NGC 4945} and Circinus are the best in the southern sky. We analyzed the 20\,MeV -- 1\,TeV spectra of the three Seyfert galaxies, \mbox{NGC 4151}, \mbox{NGC 4945} and Circinus, using 14.4\,yr of the {\it Fermi} LAT data. We derived ULs on gamma-ray emission from \mbox{NGC 4151}, which gives strong constraints on the models of high-energy neutrino emission from this source. The recent hint of a neutrino excess toward \mbox{NGC 4151}, reported by the IceCube Collaboration~\citep{IceCube:2023nai,TAUP2023}, appears to be consistent with the magnetically powered corona model. Future observations of \mbox{NGC 4151} and other Compton-thin AGNs with next-generation neutrino detectors such as IceCube-Gen2 are crucial for discriminating the different models. 
We also found evidence of sub-GeV gamma-ray emission in the direction of \mbox{NGC 4945}. If it originates from \mbox{NGC 4945}, a component other than pionic gamma-ray emission is needed. Intriguingly, this excess is consistent with CR-induced electromagnetic cascades expected in the magnetically powered corona model, although other interpretations such as high-energy emission from jets would also be possible. Cascade signatures expected in the MeV and sub-GeV gamma-ray bands are important for future MeV gamma-ray detectors such as AMEGO-X~\citep{Caputo:2022xpx} e-ASTROGAM~\citep{e-ASTROGAM:2016bph}, and GRAMS \citep{Aramaki:2019bpi}.


\begin{acknowledgements}
We thank the Topical Workshop: \mbox{NGC 1068} as cosmic laboratory sponsored by SFB1258 and Cluster of Excellence ORIGINS, at which we started this project.
The work of K.M. is supported by the NSF grants Nos.~AST-2108466 and AST-2108467, and KAKENHI Nos.~20H01901 and s20H05852. C.M.K.'s research was supported by an appointment to the NASA Postdoctoral Program at NASA Goddard Space Flight Center, administered by Oak Ridge Associated Universities under contract with NASA. S.S.K. acknowledges the support by KAKENHI Nos.~22K14028, 21H04487, and 23H04899 and the Tohoku Initiative for Fostering Global Researchers for Interdisciplinary Sciences (TI-FRIS) of MEXT’s Strategic Professional Development Program for Young Researchers. This work was supported by the European Research Council, ERC Starting grant MessMapp, S.B. Principal Investigator, under contract No.~949555.

The \textit{Fermi} LAT Collaboration acknowledges generous ongoing support from a number of agencies and institutes that have supported both the development and the operation of the LAT as well as scientific data analysis. These include the National Aeronautics and Space Administration and the Department of Energy in the United States, the Commissariat \`a l'Energie Atomique and the Centre National de la Recherche Scientifique / Institut National de Physique Nucl\'eaire et de Physique des Particules in France, the Agenzia Spaziale Italiana and the Istituto Nazionale di Fisica Nucleare in Italy, the Ministry of Education, Culture, Sports, Science and Technology (MEXT), High Energy Accelerator Research Organization (KEK) and Japan Aerospace Exploration Agency (JAXA) in Japan, and the K.~A.~Wallenberg Foundation, the Swedish Research Council and the Swedish National Space Board in Sweden. Additional support for science analysis during the operations phase is gratefully acknowledged from the Istituto Nazionale di Astrofisica in Italy and the Centre National d'\'Etudes Spatiales in France. This work performed in part under DOE Contract DE-AC02-76SF00515. Work at NRL is supported by NASA.

\end{acknowledgements}



\appendix
\section{Details of Gamma-Ray Data Analyses}
\label{sec:appendix}
\subsection{NGC 4151}
\mbox{NGC 4151} is a nearby Seyfert 1.5 galaxy, located at a distance of $d=15.8$~Mpc \citep{2020ApJ...902...26Y}. X-ray observations of \mbox{NGC 4151} indicate that it contains an ultra-fast outflow (UFO)~\citep{tombesi2010evidence,tombesi2012evidence}. Evidence for high-energy gamma-ray emission from UFOs, including \mbox{NGC 4151}, was first reported in~\citet{2021ApJ...921..144A}, based on \textit{Fermi} LAT observations. In that work, a stacking analysis was performed using a small sub-sample of the UFO population, resulting in a $5.1\sigma$ detection. 
Although \mbox{NGC 4151} itself was below the LAT detection threshold, it had an individual significance of 4.2$\sigma$, which was the highest of the sample. It should be noted that when excluding \mbox{NGC 4151} from the UFO sample, the population was still detected at the level of 3.5$\sigma$, with similar spectral properties. This showed that the signal was not due to \mbox{NGC 4151} alone. It should also be noted that in the stacking analysis, the source locations are fixed at their optical positions.  

Following the detection of the sub-threshold UFO sample, the individual detection of \mbox{NGC 4151} was recently reported in \citet{Peretti:2023crf}. 
Importantly, \mbox{NGC 4151} is located at a distance of only 0.18$^\circ$ from the fourth most significant hot spot in the search for northern neutrino point sources \citep{IceCube2022NGC1068}, and the gamma-ray detection could potentially have significant implications for the nature of the IceCube neutrinos. 

The LAT observations of \mbox{NGC 4151} are complicated by the presence of two nearby BL Lacs. In fact, one of these sources (\mbox{4FGL J1211.6+3901}, located 0.43$^\circ$ away) was carefully analyzed in \citet{Peretti:2023crf}, and it was determined that the detection of \mbox{NGC 4151} was indeed robust against the presence of this nearby gamma-ray bright blazar. The point source model in \citet{Peretti:2023crf} was originally based on the 4FGL-DR2 catalog; however, in the most recent catalog (4FGL-DR3) there is now another source (\mbox{4FGL 1210.3+3928}) that is located even closer, at a separation of 0.08$^\circ$, associated with the BL Lac, \mbox{1E 1207.9+3945}. It therefore seems that the gamma-ray emission thought to be coming from \mbox{NGC 4151} is in fact dominated by a nearby blazar.

\begin{figure*}[tb]
\includegraphics[width=0.33\linewidth]{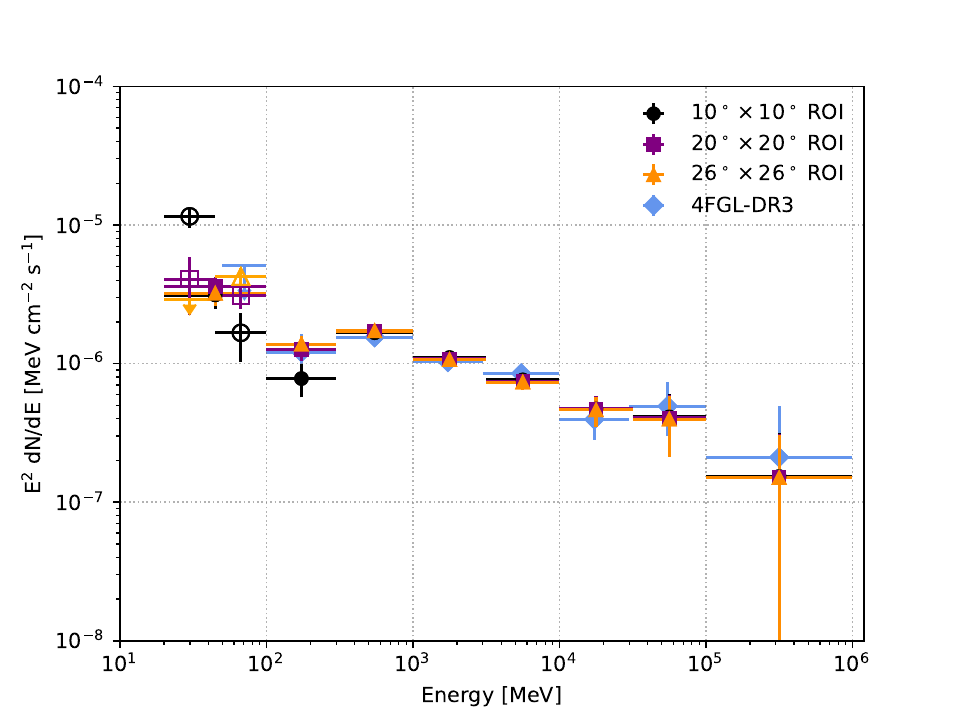}
\includegraphics[width=0.33\linewidth]{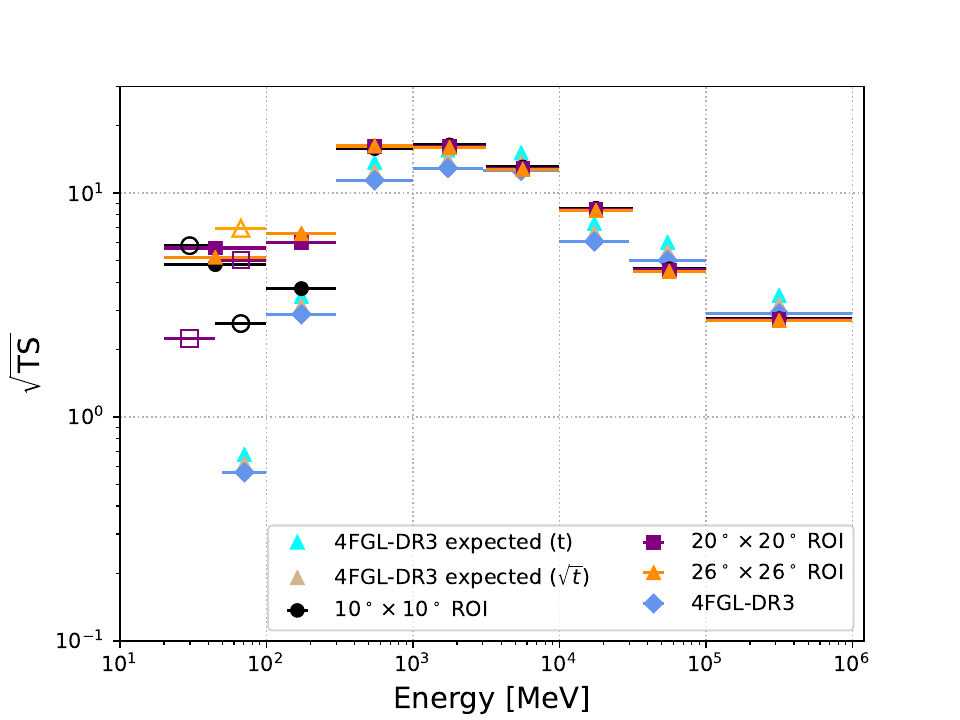}
\includegraphics[width=0.33\linewidth]{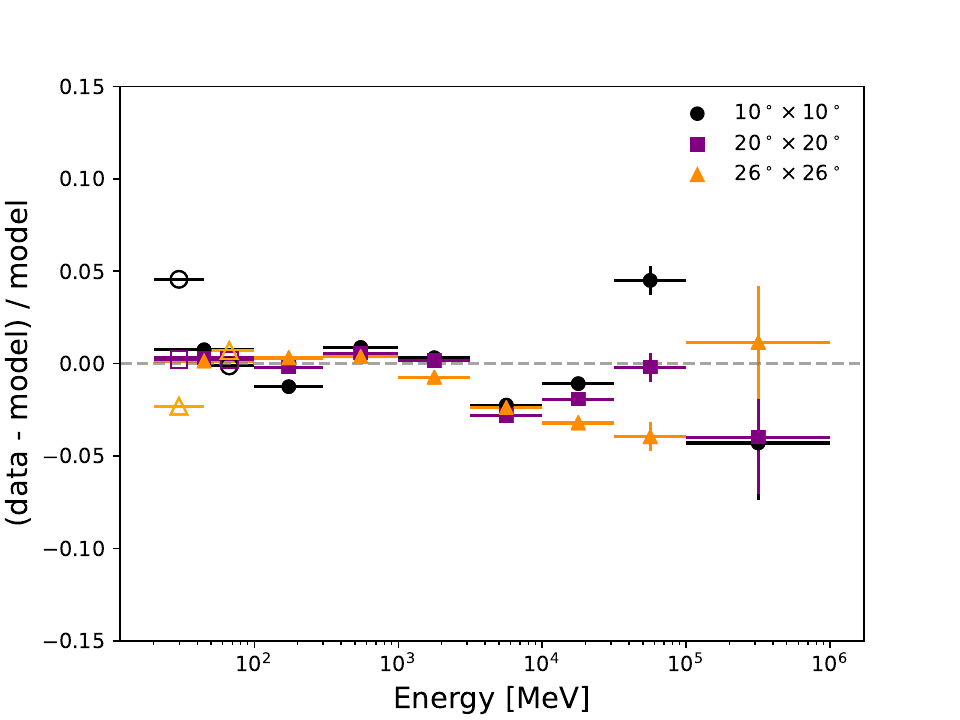}
\caption{\mbox{NGC 4945} results for three different ROI sizes, as specified in the legends. Results from the 4FGL-DR3 are also overlaid. The left panel shows the SED, the middle panel shows the corresponding significance ($\sqrt{TS}$), and the right panel shows the count residuals. In addition to the nominal binning, the first energy bin is divided into two smaller bins, shown with open markers corresponding to the ROI size. This is done to make a more direct comparison to the 4FGL-DR3, as discussed in the text. In the middle panel, we indicate the predicted significance based on the additional exposure time for both the noise-limited case ($\sqrt{t}$) and the signal-limited case ($t$). The error bars on the data points are 1$\sigma$, and the ULs are plotted for bins with TS$<$4, and shown at the $95\%$ C.L.}
\label{fig:NGC4945}
\end{figure*}
\begin{figure*}[tb]
\includegraphics[width=0.33\linewidth]{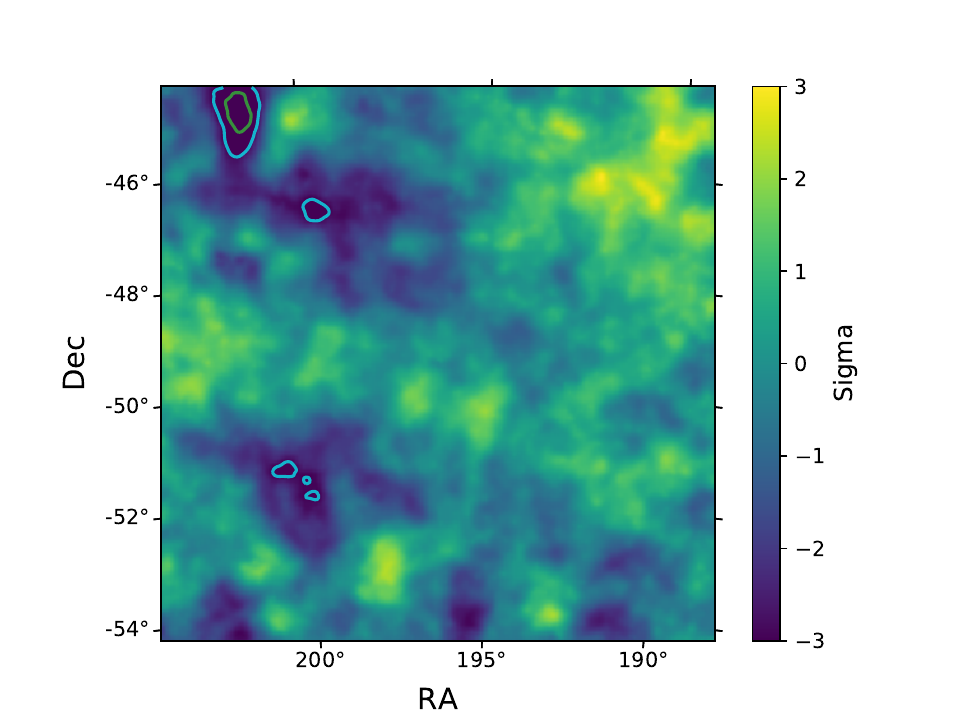}
\includegraphics[width=0.33\linewidth]{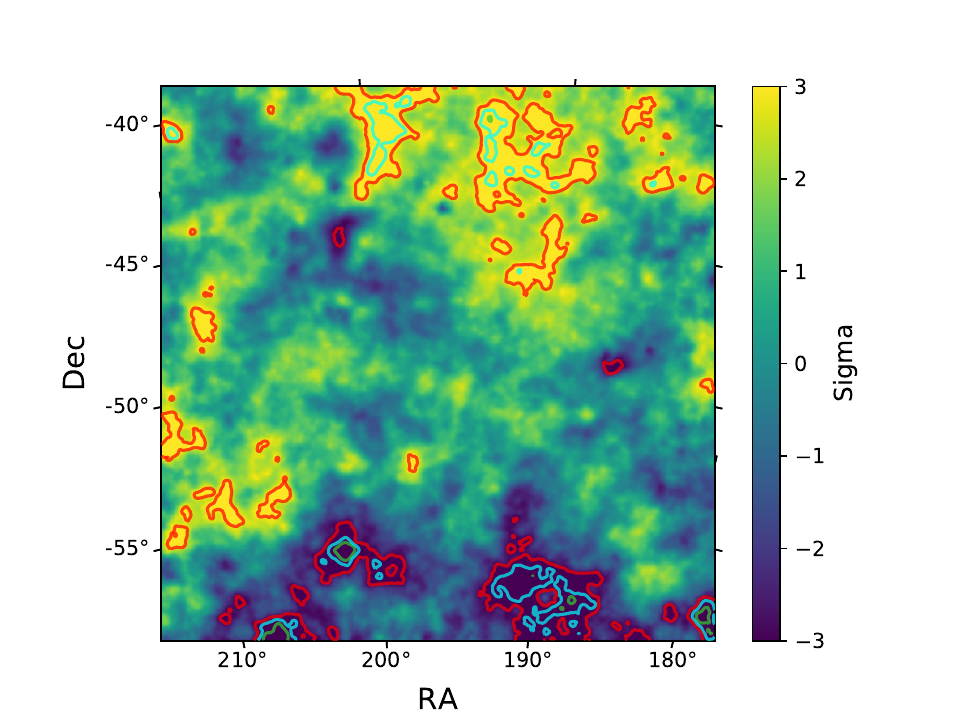}
\includegraphics[width=0.33\linewidth]{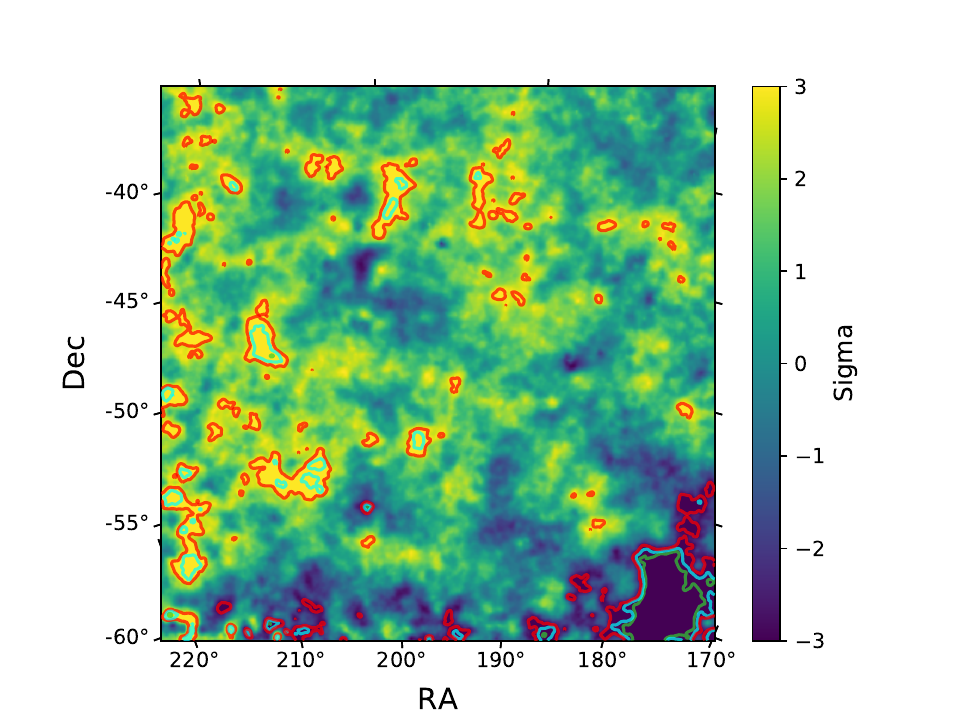}
\includegraphics[width=0.33\linewidth]{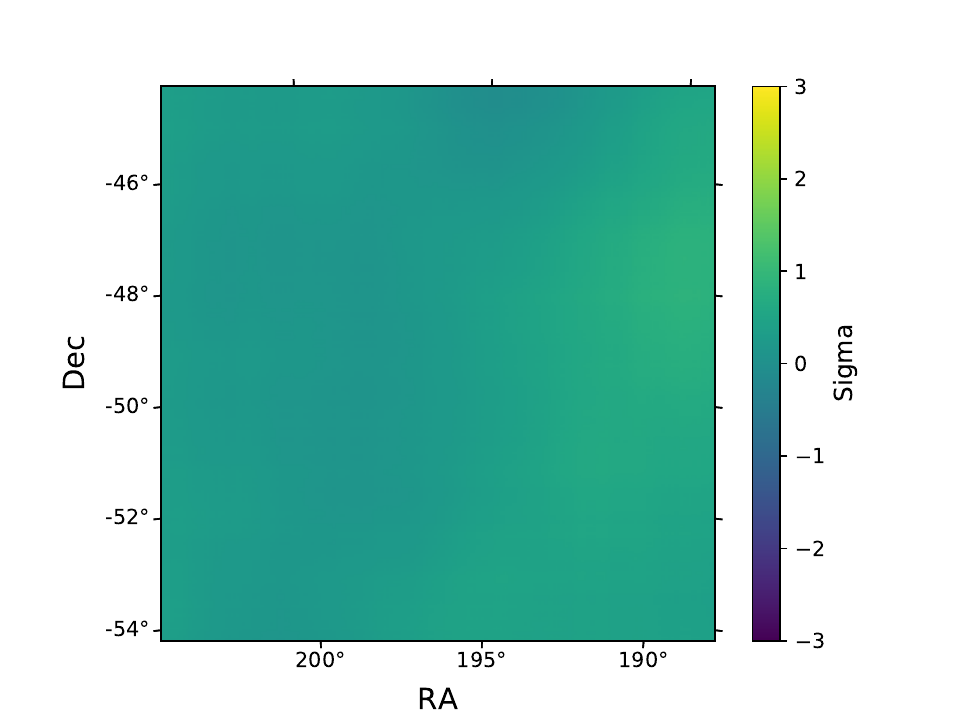}
\includegraphics[width=0.33\linewidth]{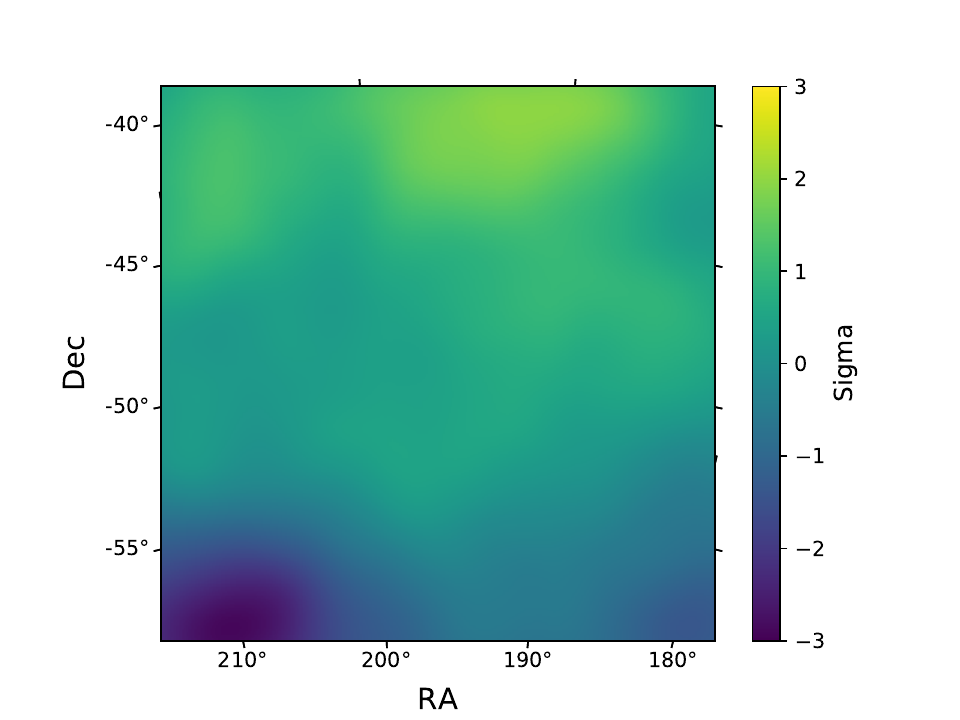}
\includegraphics[width=0.33\linewidth]{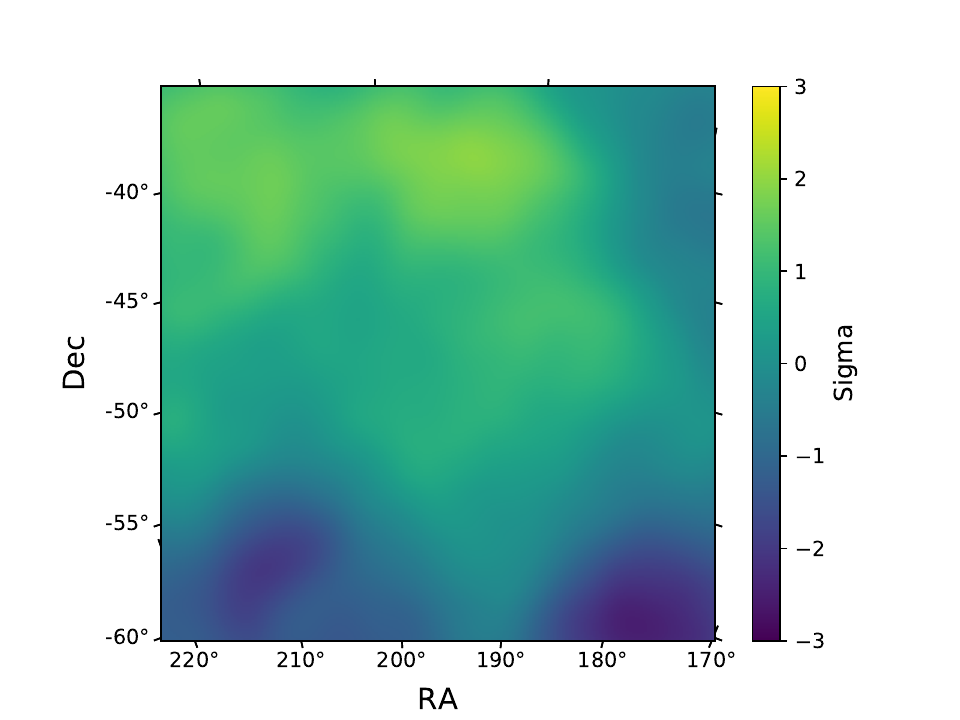}
\caption{Spatial residuals for the full energy range (top row) and first energy bin (bottom row). From left to right, the columns are $10^\circ$, $20^\circ$, and $26^\circ$ ROIs, respectively. Red, cyan, and green contours are shown at the $\pm 3\sigma, \pm 4\sigma, \mathrm{and} \pm 5 \sigma$ levels.}
\label{fig:NGC4945_II}
\end{figure*}
Based on the localization results (presented in Section~\ref{sec:data_analysis}), we keep the nearby source (\mbox{4FGL 1210.3+3928}) in the model. We add another source at the optical center of \mbox{NGC 4151} and refit the data. Unsurprisingly, \mbox{NGC 4151} is no longer detected (TS $\approx$ 0), and so we calculate ULs, shown in the left panel of Figure~\ref{fig:AGN1}, with corresponding data provided in Table~\ref{tab:NGC4151data}. The bins have TS=0, and so we calculate the ULs using a Bayesian approach (as opposed to a frequentist approach). Specifically, we use the calc\_int method from the IntegralUpperLimit class, available in the Fermi Science Tools. This method calculates an integral UL by integrating the likelihood function up to a point which contains a given fraction of the total probability. We verified that the field is well-modeled, with good data-model agreement seen in the fractional count residuals and spatial residuals. We note that in a revised version of their work, \citet{Peretti:2023crf} exclude on the basis of SED modeling that \mbox{1E 1207.9+3945} contributes to the emission of the source they associate to \mbox{NGC 4151}. In our work (and in the upcoming 4FGL-DR4) we clearly find that the  position of the gamma-ray source \mbox{4FGL 1210.3+3928} is compatible with \mbox{1E 1207.9+394}.
\begin{deluxetable*}{lcccc}
\tablecaption{NGC 4151 SED Data \label{tab:NGC4151data}}
\tablehead{
\colhead{Energy} & \colhead{Energy Low} & \colhead{Energy High} & \colhead{95\% Flux UL ($\times10^{-7}$)} & \colhead{TS} \\
\colhead{[GeV]} & \colhead{[GeV]} & \colhead{[GeV]} & \colhead{[$\mathrm{MeV \ cm^{-2} \ s^{-1}}$]} & \colhead{}
}
\startdata
$0.0447$ & $0.02$    & $0.1$  & 10.3 & 0.0 \\
$0.173$  & $0.1$    & $0.3$  & 2.6 & 0.0 \\
$0.548$  & $0.3$    & $1.0$  & 1.1 & 0.0 \\
$1.78$   & $1.0$    & $3.16$ & 0.7 & 0.0 \\
$5.62$   & $3.16$   & $10$  & 1.2 & 0.0 \\
$17.8$   & $10$     & $31.6$ & 0.8 & 0.0 \\
$56.2$   & $31.6$   & $100$  & 3.0 & 0.0 \\
$316.0$  & $100$    & $1000$  & 5.3 & 0.0 \\
\enddata
\end{deluxetable*}

\subsection{NGC 4945}
We calculate the SED for \mbox{NGC 4945}, located at the distance $d=3.6$~Mpc \citep{Monachesi:2015qla}, using the same fitting approach as was used for \mbox{NGC 4151} (in this case there is no need to relocalize). Results for this are shown in the middle panel of Figure~\ref{fig:AGN1}, and the corresponding data are provided in Table~\ref{tab:NGC4945data}. The source is detected with a TS=830. Overall, the SED is consistent with the values from the 4FGL-DR3. However, compared to the 4FGL, the two lowest-energy bins now have significant signals (TS$>$9), as opposed to ULs. For the second energy bin, this is quite reasonable given the additional exposure. Interestingly, extending the lowest-energy bin down to \mbox{20~MeV} also results in a significant signal. 

The low-energy detection of \mbox{NGC 4945} has important implications for the corona model, and so we perform additional tests to check the robustness of the emission. The 68\% containment angle at \mbox{20~MeV} for PSF3 events is $\sim 13^\circ$, and the 95\% containment angle is $\sim 30^\circ$. However, our ROI size is only $10^\circ \times 10^\circ$, and the model includes all sources within $15^\circ \times 15^\circ$. Although the ROI is smaller than the PSF at \mbox{20~MeV}, it should be sufficient for the analysis, given the following consideration. The lowest-energy bin spans the energy range \mbox{$20-100$~MeV}, and below \mbox{100~MeV}, the effective area quickly falls off. Specifically, the (total) effective area at \mbox{100~MeV} is $\sim 4.4\times$ higher compared to \mbox{20~MeV}. More of the events will therefore be coming towards the upper part of the bin, where the 68\% containment angle is $\sim 3.3^\circ$. Moreover, for a Gaussian PSF, the distribution peaks towards the center. On the other hand, the behaviour of the LAT PSF below \mbox{50~MeV} has not been well characterized. We therefore test the robustness of the emission with respect to the ROI size.

We repeat the analysis for \mbox{NGC 4945} using two different regions, motivated by the approximate $68\%$ and $95\%$ containment radii. For one test we use a $20^\circ \times 20^\circ$ ROI, and include all sources within a $40^\circ \times 40^\circ$ region. For another test, we use a $26^\circ \times 26^\circ$ ROI and include all sources within a $60^\circ \times 60^\circ$ region. Note that for the latter test, the data cover the entire $68\%$ containment region, and the model sources cover the entire $95\%$ containment region. Figure~\ref{fig:NGC4945} shows the resulting SEDs, significance ($\sqrt{\mathrm{TS}}$), and fractional count residuals for all three fits. Additionally, we overlay the SED and corresponding TS from the 4FGL-DR3. We find consistent results for all three fit variations. The flux in the first bin is compatible with the $95\%$ ULs from the 4FGL-DR3. The flux for the second energy bin is slightly lower for the $10^\circ$ ROI, but it is still consistent with the other fits within uncertainties. The fractional count residuals show that the data are well modeled, with agreement better than $5\%$ for the full energy range. The data are specifically well modeled in the first two energy bins. The spatial residuals are shown in Figure~\ref{fig:NGC4945_II} (calculated with the fermipy tool residmap) for both the entire energy range (top row) and the first energy bin (bottom row). For the full energy range, the $10^\circ$ ROI is very well modeled, whereas the larger ROIs show some excesses/deficits towards the edges of the field. For the first energy bin, all fit variations show good data-model agreement.   

In order to better understand the reason for the discrepancy in the first energy bin with respect to the 4FGL-DR3, we divide it into two smaller bins to match the 4FGL binning, and recalculate the SED. The lower part of the bin spans the energy range \mbox{$20 - 45$~MeV}, and the upper part of the bin spans the energy range \mbox{$45 - 100$~MeV}. We make the split at \mbox{45~MeV} (compared to \mbox{50~MeV} for the lower edge of the 4FGL bin), as required by our initial energy binning. This test is performed for all three ROI variations, and the results are shown in Figure~\ref{fig:NGC4945} with open markers, corresponding to the respective ROI size.

For our nominal ROI size of $10^\circ$, we find a significance ($\sqrt{\mathrm{TS}}$) of 5.8 and 2.6 for the two bins. The significance of the upper part of the bin is now reasonable given that of the 4FGL, and the flux is also compatible with the 4FGL ULs. Additionally, we find a significant signal in the lower portion of the bin. However, the corresponding count residuals in the lower portion of the bin are slightly elevated at the level of $\sim 5\%$. This can plausibly be attributed to some mismodeling of either the isotropic emission or the Galactic diffuse emission (which at these energies is dominated by inverse-Compton radiation). Note that the isotropic model uses a power-law extrapolation below \mbox{34~MeV}, and the Galactic diffuse model uses a power-law extrapolation below \mbox{50~MeV}. For the $20^\circ$ ROI, the flux of the two smaller bins is found to be consistent with that of the combined bin, and likewise the count residuals are close to 0. For the $26^\circ$ ROI, the upper portion of the bin is also compatible with the combined bin; however, the lower portion of the bin is over-modeled by $\sim 3\%$, and the SED calculation results in only an UL, albeit at the same level as the observed signal. 

As another test, we have made the SED calculation using a spectral index of 3.0 (instead of the commonly used value of 2.0), consistent with our model predictions, and also freeing all point sources within $5^\circ$, having a TS$>$16. This test was run for the $20^\circ$ ROI. We have found that the low-energy bin ($E_\gamma<100$~MeV) is still detected with a similar flux having a TS of 19.0. Finally, we have tested the SED calculation when freeing all sources within $7^\circ$ having a spectral index $>2.5$. In this case, we have found consistent results for the low-energy bin having a TS of 27.5.

Overall, these tests show that the emission between \mbox{$20-100$~MeV} appears to be robust, although there is clearly a systematic uncertainty in the measured flux of a factor of $\sim2-3$. However, this result should still be interpreted with caution, due to the generally poor performance of the LAT below \mbox{50~MeV}.

\begin{deluxetable*}{lcccccc}
\tablecaption{NGC 4945 SED Data \label{tab:NGC4945data}}
\tablehead{
\colhead{Energy} & \colhead{Energy Low} & \colhead{Energy High} & \colhead{Flux ($\times10^{-7}$)} & \colhead{1$\sigma$ Flux Error ($\times10^{-7}$)} & \colhead{95\% Flux U.L ($\times10^{-7}$)} & \colhead{TS} \\
\colhead{[GeV]} & \colhead{[GeV]} & \colhead{[GeV]} & \colhead{[$\mathrm{MeV \ cm^{-2} \ s^{-1}}$]} & \colhead{[$\mathrm{MeV \ cm^{-2} \ s^{-1}}$]} & \colhead{[$\mathrm{MeV \ cm^{-2} \ s^{-1}}$]} & \colhead{}
}
\startdata
$0.0447$  & $0.02$     & $0.1$  & 31.3  & 6.5 & 42.1 & 23.0 \\
$0.173$   & $0.1$     & $0.3$  & 7.8   & 2.1 & 11.3 & 14.0 \\
$0.548$   & $0.3$     & $1.0$  & 16.7  & 1.2 & 18.7 & 251 \\
$1.78$    & $1.0$     & $3.16$ & 11.1  & 0.9 & 12.6 & 270 \\
$5.62$    & $3.16$    & $10$  & 7.6   & 1.0 & 9.3 & 171 \\
$17.8$    & $10$     & $31.6$ & 4.7   & 1.1 & 6.9 & 73.3 \\
$56.2$    & $31.6$    & $100$  & 4.1   & 1.9 & 7.9 & 21.1 \\
$316.0$   & $100$     & $1000$  & 1.5   & 1.6 & 5.8 & 7.6 \\
\enddata
\end{deluxetable*}

\subsection{Circinus}
We also calculate the SED for Circinus, located at $d=4.2$ Mpc \citep{1977A&A....55..445F}; the results are shown in the right panel of Figure~\ref{fig:AGN1}, and the corresponding data are provided in Table~\ref{tab:circdata}. The source is detected with a TS=130. Overall, the SED is consistent with the values from the 4FGL-DR3. The fractional count residuals are very good, being close to 0 for all energies.

\begin{deluxetable*}{lcccccc}
\tablecaption{Circinus SED Data \label{tab:circdata}}
\tablehead{
\colhead{Energy} & \colhead{Energy Low} & \colhead{Energy High} & \colhead{Flux ($\times10^{-7}$)} & \colhead{1$\sigma$ Flux Error ($\times10^{-7}$)} & \colhead{95\% Flux U.L ($\times10^{-7}$)} & \colhead{TS} \\
\colhead{[GeV]} & \colhead{[GeV]} & \colhead{[GeV]} & \colhead{[$\mathrm{MeV \ cm^{-2} \ s^{-1}}$]} & \colhead{[$\mathrm{MeV \ cm^{-2} \ s^{-1}}$]} & \colhead{[$\mathrm{MeV \ cm^{-2} \ s^{-1}}$]} & \colhead{}
}
\startdata
$0.0447$  & $0.02$     & $0.1$  & 22.8  & 13.9 & 45.8 & 2.7 \\
$0.173$   & $0.1$     & $0.3$  & 6.5   & 4.0 & 13.1 & 2.7 \\
$0.548$   & $0.3$     & $1.0$  & 6.8  & 1.6 & 9.5 & 18.6 \\
$1.78$    & $1.0$     & $3.16$ & 5.2  & 1.0 & 6.9 & 32.5 \\
$5.62$    & $3.16$    & $10$  & 4.3   & 0.9 & 5.8 & 40.6 \\
$17.8$    & $10$     & $31.6$ & 2.6   & 1.0 & 4.5 & 15.6 \\
$56.2$    & $31.6$    & $100$  & 2.7   & 1.5 & 5.8 & 14.7 \\
$316.0$   & $100$     & $1000$  & 2.3   & 2.0 & 7.0 & 3.5 \\
\enddata
\end{deluxetable*}

\bibliographystyle{aasjournal}
\bibliography{kmurase}

\end{document}